# Spontaneous CP violation and CPT violation

Yu Kun Qian

*Physics School of Peking University,Beijing 100871,P.R.China*

At first we give a little formalism to show some features of spontaneous CP violation theory. Then we give a convincing argument show that Cronin etc's experiment is a evidence of CPT violation and spontaneous CP violation is absolutely necessary. Final we discuss some possible CPT violation mechanism.

At first we present a little formalism on spontaneous CP violation[1]. Let $H^0$ be a neutral real Higgs field, to have it to be CP negative and T negative we have to make a pseudoscalar coupling

$$\text{L} = ig\overline{\Psi}\gamma^5\Psi H^0$$

Where $g$ is real and the coupling is supposed to be C,P,T invariant. Then we have $CPH^0(CP)^{-1} = -H^0$ and $TH^0T^{-1} = -H^0$. If $H^0$ has a nonvanishing vacuum expectation value

$$\langle 0|H^0|0\rangle = v$$

Then we have

$$\langle 0|(CP)^{-1}H^0CP|0\rangle = -v$$

So $CP|0\rangle$ no long be equal to $|0\rangle$ but a degenerated vacuum, but $CPT|0\rangle$ still equals to $|0\rangle$, this is the key point of spontaneous CP violation theory. Now a CP invariant Hamiltonian will not guarantee the physical processes to be CP invariant. We can introduce spontaneous CP violation to W-S model. In W-S model we have Higgs field

$$H = \begin{pmatrix} H^+ \\ H^0 \end{pmatrix}$$

Where $H^0$ is a complex field but we can define 2 real fields

$$H_1^0 = \frac{H^0 + H^{0+}}{\sqrt{2}}$$

$$H_2^0 = \frac{H^0 - H^{0+}}{\sqrt{2}i}$$

The $H_2^0$ is a Goldstone boson we can put it aside. The CP transformation property of $H_1^0$ is based on the Ukawa coupling. We can make the Ukawa coupling a scalar coupling or a pseudoscalar coupling. In the standard model usually we made the Ukawa coupling a scalar coupling. In this case we have

$$CPH_1^0(CP)^{-1} = H_1^0$$

Then we don't have spontaneous CP violation. In the next step we will demonstrate how to make the Ukawa coupling a pseudoscalar coupling. We demonstrate it with the Ukawa coupling between $u, c$ quarks as a example. The Ukawa coupling can be written as

$$g_u \begin{pmatrix} \bar{u} & \bar{d} \end{pmatrix}_L u_R \begin{pmatrix} H^{0+} \\ -H^- \end{pmatrix} + g'_u \bar{u}_R \begin{pmatrix} H^0 & -H^+ \end{pmatrix} \begin{pmatrix} u \\ d \end{pmatrix}_L$$

$$+ g_c \begin{pmatrix} \bar{c} & \bar{s} \end{pmatrix}_L c_R \begin{pmatrix} H^{0+} \\ -H^- \end{pmatrix} + g'_c \bar{c}_R \begin{pmatrix} H^0 & -H^+ \end{pmatrix} \begin{pmatrix} c \\ s \end{pmatrix}_L$$

$$+ g_{uc} \begin{pmatrix} \bar{u} & \bar{d} \end{pmatrix}_L c_R \begin{pmatrix} H^{0+} \\ -H^- \end{pmatrix} + g'_{uc} \bar{c}_R \begin{pmatrix} H^0 & -H^+ \end{pmatrix} \begin{pmatrix} u \\ d \end{pmatrix}_L$$

$$+ g_{cu} \begin{pmatrix} \bar{c} & \bar{s} \end{pmatrix}_L u_R \begin{pmatrix} H^{0+} \\ -H^- \end{pmatrix} + g'_{cu} \bar{u}_R \begin{pmatrix} H^0 & -H^+ \end{pmatrix} \begin{pmatrix} c \\ s \end{pmatrix}_L$$

The hermiticity require

$$g'_u = g_u^*, g'_c = g_c^*, g'_{uc} = g_{uc}^*, g'_{cu} = g_{cu}^*$$

Throw away the Goldstone particles part and if we want the coupling to be pseudoscalar we have to let $g_u = ig_1, g_c = ig_2, g_{uc} = ig_{12}, g_{cu} = ig_{21}, g_{12} = g_{21}$ where $g_1, g_2, g_{12}$ are all real. Then we right have the pseudoscalar coupling

$$\frac{ig_1}{\sqrt{2}} \bar{u}\gamma^5 u H_1^0 + \frac{ig_2}{\sqrt{2}} \bar{c}\gamma^5 c H_1^0 + \frac{ig_{12}}{\sqrt{2}} \bar{u}\gamma^5 c H_1^0 + \frac{ig_{21}}{\sqrt{2}} \bar{c}\gamma^5 u H_1^0$$

In this case $CPH_1^0(CP)^{-1} = -H_1^0$ then the nonvanishing vacuum expectation of $H_1^0$ means spontaneous CP violation

$$CP|0\rangle \neq |0\rangle$$

Moreover the terms with the vacuum expectation value are not normal mass terms, the fermion masses have to be generated by renormalization let the terms as some 2 point interaction. For demonstration we consider a single fermion, we calculate its "complete" propagator, we have it to be a sum

$$\frac{i}{\slashed{p}} + \frac{i}{\slashed{p}} gv\gamma^5 \frac{i}{\slashed{p}} + \frac{i}{\slashed{p}} gv\gamma^5 \frac{i}{\slashed{p}} gv\gamma^5 \frac{i}{\slashed{p}} + \frac{i}{\slashed{p}} gv\gamma^5 \frac{i}{\slashed{p}} gv\gamma^5 \frac{i}{\slashed{p}} gv\gamma^5 \frac{i}{\slashed{p}} + \ldots$$

$$= \frac{i}{\slashed{p}}(1 + \frac{g^2v^2}{p^2} + \frac{g^2v^2}{p^2}\frac{g^2v^2}{p^2} + \ldots) + \frac{gv\gamma^5}{P^2}(1 + \frac{g^2v^2}{p^2} + \frac{g^2v^2}{p^2}\frac{g^2v^2}{p^2} + \ldots)$$

$$= \frac{i\slashed{p}}{p^2}\frac{1}{1 - \frac{g^2v^2}{p^2}} + \frac{gv\gamma^5}{p^2}\frac{1}{1 - \frac{g^2v^2}{p^2}}$$

$$= \frac{i\slashed{p} + gv\gamma^5}{p^2 - g^2v^2}$$

It has a pole at $gv$, so the effective mass of the fermion is $gv$.

Another issue has to be pointed out is that if the Ukawa coupling is written as a scalar coupling we have a mass matrix to be complex hermitian, but if the Ukawa coupling is written as a pseudoscalar coupling we have a "mass" matrix to be imaginary symmetric then the CKM matrix can readily be taken to be a real orthogonal matrix. By the way we recall in the usual W-S model we have

$$u_i = V_{ij} u_{0j}$$

Where $u_i$ is i-th physical up quark, $u_{0j}$ is j-th original up quark, $V_{ij}$ is the CKM matrix element, general speaking is complex. Now problem is emerging: if we define the charge conjugation of $u_i$ as $u_i^C = i\gamma^2 u_i^{+T}$ and the charge conjugation of $u_{0j}$ as $u_{0j}^C = i\gamma^2 u_{0j}^{+T}$ we will find the two definitions are conflicting. First since charge conjugation is a unitary transformation then we should have

$$u_i^C = V_{ij} u_{0j}^C$$

Second since $u_i^{+T} = V_{ij}^* u_{0j}^{+T}$ then we have

$$u_i^C = V_{ij}^* u_{0j}^C$$

So they are conflicting and we know the two charge conjugation definitions cannot be made at same time. Then it perhaps is a natural choice let the CKM matrix to be real and all the CP violation effects are solely due to spontaneous CP violation.

Next we will try to reexamine what does the Cronin etc's experiment[2] means. We begin with considering C,P,T transformation of pions. C,P, T transformations of pions is based on the pseudoscalar coupling between nucleons and pions

$$\mathcal{L} = ig\bar{N}\gamma_5 \sigma N \phi$$

where $g$ is real and $i$ is required by hermiticity. The coupling is a strong interaction and supposed to be C,P invariant and time reversal invariant. Then we have

$$C\pi^+(x,t)C^{-1} = \pi^-(x,t)$$
$$C\pi^-(x,t)C^{-1} = \pi^+(x,t)$$
$$C\pi^0(x,t)C^{-1} = \pi^0(x,t)$$

For creation and annihilation operators we have

$$U_C a_{\pi^+ p} U_C^{-1} = a_{\pi^- p}, \quad U_C a_{\pi^+ p}^+ U_C^{-1} = a_{\pi^- p}^+$$
$$U_C a_{\pi^- p} U_C^{-1} = a_{\pi^+ p}, \quad U_C a_{\pi^- p}^+ U_C^{-1} = a_{\pi^+ p}^+$$
$$U_C a_{\pi^0 p} U_C^{-1} = a_{\pi^0 p}, \quad U_C a_{\pi^0 p}^+ U_C^{-1} = a_{\pi^0 p}^+$$

And

$$P\pi_i(x,t)P^{-1} = -\pi_i(-x,t)$$

For creation and annihilation operators we have

$$U_P a_{ip} U_P^{-1} = -a_{i-p}, \quad U_P a_{ip}^+ U_P^{-1} = -a_{i-p}^+$$

And

$$T\pi^+(x,t)T^{-1} = -\pi^+(x,-t)$$

$$T\pi^-(x,t)T^{-1} = -\pi^-(x,-t)$$

$$T\pi^0(x,t)T^{-1} = -\pi^0(x,-t)$$

For creation and annihilation operators we have

$$U_T a_{ip} U_T^{-1} = -a_{i-p}, \quad U_T a_{ip}^+ U_T^{-1} = -a_{i-p}^+$$

where subscript $i$ is for $\pi^\pm$ or $\pi^0$.

The transformations of 2 pions S states are as following

$$|2\pi_0\rangle = \int d^3p f_1(p^2) a_{\pi_0 p}^+ a_{\pi_0 -p}^+ |0\rangle$$

$$|\pi^+\pi^-\rangle = \int d^3p f_2(p^2) a_{\pi^+ p}^+ a_{\pi^- -p}^+ |0\rangle$$

$$CP|2\pi_0\rangle = \int d^3p f_1(p^2) a_{\pi_0 p}^+ a_{\pi_0 -p}^+ CP|0\rangle$$

$$TCP|2\pi_0\rangle = |2\pi_0\rangle$$

$$CP|\pi^+\pi^-\rangle = \int d^3p f_2(p^2) a_{\pi^+ p}^+ a_{\pi^- -p}^+ CP|0\rangle$$

$$TCP|\pi^+\pi^-\rangle = |\pi^+\pi^-\rangle$$

When there is no spontaneous CP violation $CP|0\rangle = |0\rangle$ then 2 pions S states are CP eigenstates with positive eigenvalue, otherwise they are not. But in either cases they are CPT eigenstates with positive eigenvalue.

Next we look at the $K^0, \overline{K}^0$ states. In the history sometime people define $\overline{K}^0$ as $K^0$'s CP conjugate, sometime people define $\overline{K}^0$ as $K^0$'s CPT conjugate. We have to point out that the two definitions cannot be made arbitrarily for that everytime when you make a definition you have to adjust the phase factor between $|K^0\rangle$ and $|\overline{K}^0\rangle$. So two definitions may be conflicting. To make a definite C,P,T transformation property we extend the pseudoscalar coupling between nucleons and pions to a pseudoscalar coupling between eightfold barions and eightfold pseudoscalar mesons

$$\mathcal{L} = ig_1 \overline{B}_{ab} \gamma_5 B_{bc} \phi_{ca} + ig_2 \overline{B}_{ab} \gamma_5 B_{ca} \phi_{bc}$$

Where

$$B_{ab} = \begin{pmatrix} \frac{\Sigma^0}{\sqrt{2}} + \frac{\Lambda^0}{\sqrt{6}} & \Sigma^+ & p \\ \Sigma^- & -\frac{\Sigma^0}{\sqrt{2}} + \frac{\Lambda^0}{\sqrt{6}} & n \\ \Xi^- & \Xi^0 & -\frac{2\Lambda^0}{\sqrt{6}} \end{pmatrix}$$

$$\bar{B}_{ab} = \begin{pmatrix} \frac{\bar{\Sigma}^0}{\sqrt{2}} + \frac{\bar{\Lambda}^0}{\sqrt{6}} & \bar{\Sigma}^- & \bar{\Xi}^- \\ \bar{\Sigma}^+ & -\frac{\bar{\Sigma}^0}{\sqrt{2}} + \frac{\bar{\Lambda}^0}{\sqrt{6}} & \bar{\Xi}^0 \\ \bar{p} & \bar{n} & -\frac{2\bar{\Lambda}^0}{\sqrt{6}} \end{pmatrix}$$

$$\phi_{ab} = \begin{pmatrix} \frac{\pi^0}{\sqrt{2}} + \frac{\eta}{\sqrt{6}} & \pi^+ & K^+ \\ \pi^- & \frac{-\pi^0}{\sqrt{2}} + \frac{\eta}{\sqrt{6}} & K^0 \\ K^- & \bar{K}^0 & \frac{-2\eta}{\sqrt{6}} \end{pmatrix}$$

$g_1, g_2$ are real and $i$ is for hermiticity and $\bar{K}^0 = (K^0)^+$. The coupling is also a strong interaction and should to be supposed C,P,T invariant. Then we easy to find as above

$$Ca^+_{K^0 p} C^{-1} = a^+_{\bar{K}^0 p}$$
$$Ca^+_{\bar{K}^0 p} C^{-1} = a^+_{K^0 p}$$
$$Pa^+_{K^0 p} P^{-1} = -a^+_{K^0 -p}$$
$$Pa^+_{\bar{K}^0 p} P^{-1} = -a^+_{\bar{K}^0 -p}$$
$$Ta^+_{K^0 p} T^{-1} = -a^+_{K^0 -p}$$
$$Ta^+_{\bar{K}^0 p} T^{-1} = -a^+_{\bar{K}^0 -p}$$

Then we have

$$CPa^+_{K^0} (CP)^{-1} = -a^+_{\bar{K}^0}$$

And also

$$CPa^+_{\bar{K}^0} (CP)^{-1} = -a^+_{K^0}$$

And

$$TCPa^+_{K^0} (TCP)^{-1} = a^+_{\bar{K}^0}$$

And also

$$TCPa^+_{\bar{K}^0} (TCP)^{-1} = a^+_{K^0}$$

Now we define

$$|K_1\rangle = \frac{1}{\sqrt{2}}(|K^0\rangle - |\bar{K}^0\rangle)$$

$$|K_2\rangle = \frac{1}{\sqrt{2}}(|K^0\rangle + |\bar{K}^0\rangle)$$

Then $|K_1\rangle$ is CP positive eigenstate $|K_2\rangle$ is CP negative eigenstate when there is no spontaneous CP violation otherwise they are not CP eigenstates. Also $|K_1\rangle$ is CPT negative eigenstate $|K_2\rangle$ is CPT positive eigenstate no matter there is or not spontaneous CP violation. We see that the CP conjugates and CPT conjugates just differ from each other by a minus sign. This minus sign is of vital importance. Now suppose there is no spontaneous CP violation then if $|K_1\rangle$ is nearly $|K_S\rangle$ it decays into 2 pions states with a larger probability but it is CPT negative so it means CPT negative state decays into 2 pions states, which are CPT positive states, with a larger probability, otherwise if $|K_2\rangle$ is nearly $|K_S\rangle$ but it is CP negative so it means CP negative state decays into 2 pions states,which are CP positive states, with a larger probability. Either cases are unreasonable. So the only possibility is we have to suggest spontaneous CP violation. Moreover now $|K_2\rangle$ is nearly $|K_S\rangle$ it is nearly a CPT eigenstate with positive eigenvalue if $|K_L\rangle$ also decays into two pions final states it is a signal of CPT violation. So Cronin etc's experiment is not a evidence of CP violation but a evidence of CPT violation. The next natural problem is what is the CPT violation mechanism. As well known there is a strong CPT theorem. But CPT theorem was raised in 50s last century that time people don't know $W^\pm, Z^0$,Higgs, gluon and ghost. By a simple checking easy to find $W^\pm, Z^0$,Higgs, gluon terms are all CPT invariant only left is ghost. Ghost is a scalar but anti-commuting so it violate the precondition of CPT theorem. Look at the QCD effective action it's easy to see the two terms concerning ghost are CPT violating .

[1]T.D.Lee,Physics Reports **9c**,No2(1974)
[2]J.H.Christenson,J.W.Cronin,V.L.Fitch and R.Turlay, Phys.Rev.Lett.**13,**138(1964)